\documentclass{amsart}
\pdfoutput=1
\normalsize
\usepackage{lipsum}
\usepackage{amsfonts}
\usepackage{graphicx}
\usepackage{float}

\usepackage{amsmath}
\usepackage{amssymb}
\usepackage{ulem}

\usepackage{tikz}
\usetikzlibrary{shapes,arrows}

\DeclareMathOperator*{\argmin}{arg\,min}

\usepackage{enumitem}
\setlist[enumerate]{leftmargin=.5in}
\setlist[itemize]{leftmargin=.5in}

\usepackage[english]{babel}

\usepackage[letterpaper,top=2cm,bottom=2cm,left=3cm,right=3cm,marginparwidth=1.75cm]{geometry}

\usepackage{amsmath}
\usepackage{graphicx}
\usepackage[colorlinks=true, allcolors=blue]{hyperref}
\usepackage{cleveref}
\usepackage{caption}
\usepackage{subcaption}

\usepackage{epstopdf}
\usepackage[ruled,lined]{algorithm2e}

\title{Multi-scale Hybridized Topic Modeling: A Pipeline for analyzing unstructured text datasets via Topic Modeling}

\author{Keyi Cheng $^{1,\dagger}$}
\address{$^1$University of California, Los Angeles}
\address{$^2$University of California, Berkeley}
\address{$^3$The University of Texas Rio Grande Valley (UTRGV)}
\address{$^{\dagger}$Equal contributors}
\author{Stefan Inzer $^{2,\dagger}$}
\author{Adrian Leung $^{1,\dagger}$}
\author{Xiaoxian Shen $^{1,\dagger}$}

\author{\\Advisors: Michael Perlmutter$^{1}$}
\author{Michael Lindstrom$^{3}$}
\author{Joyce Chew$^{1}$}
\author{Todd Presner$^{1}$}
\author{Deanna Needell$^{1}$}

\usepackage[foot]{amsaddr}

\begin{document}

\maketitle

\begin{abstract}
We propose a multi-scale hybridized topic modeling method to find hidden topics from transcribed interviews more accurately and more efficiently
than traditional topic modeling methods. Our multi-scale hybridized topic modeling method (MSHTM) approaches data at different scales and performs topic modeling in a hierarchical way utilizing first a classical method, Nonnegative Matrix Factorization, and then a transformer-based method, BERTopic. 
It harnesses the strengths of both NMF and BERTopic. Our method can help researchers and the public better extract and interpret the interview information. Additionally, it provides insights for new indexing systems based on the topic level. We then deploy our method on real-world interview transcripts and find promising results.
\end{abstract}

\section{Introduction}

 Transcribed interviews are important sources of information. They often provide accurate firsthand evidence, which is useful for analyzing human experiences and events. Since interviews usually contain a large amount of information, manually analyzing them is an arduous and impractical task. To gain a useful overview of the interviews more efficiently, a beneficial exercise is to extract and group information by relevant topics from the interviews. This helps those analyzing transcribed interviews understand both the general and detailed events mentioned in interviews and locate information within their field of interest. Traditionally, interview transcripts have been indexed topically by humans, and while nominal indexing is usually done correctly, the human resources required to index an interview by hand are costly, inefficient, and unfeasible with large-scale text datasets.

To address this problem, it becomes a computational task to extract meaningful and detailed topics from text documents through precise and efficient methods. To this end, topic modeling algorithms such as Latent Dirichlet Allocation (LDA), Non-negative Matrix Factorization (NMF), Top2Vec, and BERTopic have been developed to find hidden topics from text documents and provide the readers with a better overview of the contents and structures of the text documents, especially when the text documents are large. As noted in previous comparisons between the four popular topic modeling methods, both BERTopic and NMF have better performances over the other two methods \cite{anantharaman2019performance, egger2022topic, ray2019review}. Additionally, LDA is a probabilistic method that uses raw word counts and is based on many assumptions which NMF does not assume \cite{blei2003latent}. Between the two embedding methods, BERTopic generates more clear-cut topics than Top2Vec \cite{egger2022topic}.

Our method combines NMF and BERTopic in order to take advantage of the relative strengths of each method. NMF is useful for finding interpretable broad topics and has the flexibility to assign multiple topics to a given document. However, NMF relies on a bag-of-words representation, which does not incorporate contextual information \cite{egger2022topic}. NMF can also miss more detailed subtopics that are important for understanding groups of interviewees' experiences. Hierarchical NMF can detect subtopics, but traditional methods usually work better on a document level as short texts suffer data sparsity problems \cite{chen2019experimental, cheng2014btm}. 
BERTopic, on the other hand, is useful for finding detailed topics at the sentence level, but it produces an excessively large number of topics that become infeasible to read through \cite{egger2022topic, grootendorst2022bertopic}. Each document can only be assigned to a single topic, which is a major limitation when a given document contains several topics. BERTopic also has a token limitation, which limits its performance on large documents. Therefore, we shall combine these two approaches to perform a more comprehensive topic modeling.

In this paper, we introduce a novel hierarchical method, Multi-Scale Hybridized Topic Modeling (MSHTM), that combines NMF and BERTopic in order to efficiently discover detailed and semantically rich topics over interview transcripts. 
The paper is organized as follows:  background information is in
\ref{sec:background}, our method is overviewed in \Cref{sec:methods}, the datasets we used to test our methods are described in \Cref{sec:datasets}, the results and visualizations are in \Cref{sec:results and visualizations}, our discussions of the use cases for MSHTM are in \Cref{sec:discussions}, and the conclusions follow in
\Cref{sec:conclusions}.

\section{Background}
\label{sec:background}

In this section, we review Topic Modeling and two key topic modeling methods: NMF and the transformer-based BERTopic algorithm. 

\subsection{Overview of Topic Modeling}
Topic Modeling is the task of extracting latent topics from unstructured data and is a helpful tool for understanding texts and organizing text documents. Often, it is worthwhile to identify the underlying hierarchical relationship between topics, and various methods have been developed to learn topics at different granularities. In the rest of this section, we will discuss NMF and BERTopic in detail.

\subsection{Notation} We will follow the same set of notations in the rest of the paper. A matrix $\textbf{X}$, a vector $\textbf{x}$, and a scalar $x$. $\textbf{X}_{i,j}$ represents the $i$th row, $j$th column element of matrix $\textbf{X}$. $\begin{Vmatrix} {A} \end{Vmatrix}^{2}_{F} = \sum_{i, j} |A_{ij}|^2$ 
represents the Frobenius norm. $\mathbb{R}^{d \times n}_{\geq{0}}$ denotes non-negative real space of dimension $d \times n$.

\subsection{NMF}
\label{subsec:NMF}

NMF
is a widely used unsupervised learning method popularized by D. Lee 
and H. Seung \cite{lee1999learning}. The non-negativity constraint 
helps to decompose data in a naturally interpretable way, as opposed to eigenvector-based methods such as Singular Value Decomposition (SVD) and Principle Component Analysis (PCA).

Classical NMF processes the data matrix by decomposing it into the product of two low-rank non-negative matrices. Given a data matrix $\textbf{X} \in \mathbb{R}^{d \times n}_{\geq{0}}$, the goal is to approximate the original data matrix with two non-negative matrices: the dictionary matrix $\textbf{W} \in \mathbb{R}^{d \times r}_{\geq{0}}$ and the coding matrix $\textbf{H} \in \mathbb{R}^{r \times n}_{\geq{0}}$, where $n$ is the number of documents, $d$ is the number of words in the dictionary, and $r$ is the number of topics. This is achieved by minimizing the following loss function 
\[
(\textbf{W},\textbf{H}) = \argmin_{\textbf{W} \in \mathbb{R}^{d \times r}_{\geq{0}},\, \textbf{H} \in \mathbb{R}^{r \times n}_{\geq{0}}}\begin{Vmatrix} {\textbf{X} - \textbf{WH}} \end{Vmatrix}^{2}_{F}.
\] 
Each column of $\textbf{X}$ corresponds to a document, where each document is vectorized into term frequency-inverse document frequency (TF-IDF) representation based on the frequencies of words, see more details in \Cref{Classical NMF}. 
Each column of $\textbf{W}$ corresponds to a dictionary representation of a topic. Each column of $\textbf{H}$ corresponds to the coding for a given document in relation to all the topics found, and each row of $\textbf{H}$ corresponds to the relevance of a specific topic for all the documents. 


The success of NMF in parts-based learning comes from its non-negativity constraints. Compared to the potentially canceling combinations obtained in SVD and PCA, the computed matrices, $\textbf{W}$ and $\textbf{H}$, are naturally sparse—comprising of mostly zero entries—and hence create more expressive and straightforward interpretations of text documents, allowing NMF to be powerful in various applications including topic modeling, news classification, and image processing.

One particularly useful family of NMF variants is Hierarchical NMF, which extracts topics from the data at varying levels of granularity and attempts to capture the relationship between them. These variants include often involve
 detecting a tree structure between supertopics and subtopics \cite{kuang2013fast, grotheer-etal-2020-covid} and organizing and reducing existing topics \cite{9022678}. One top-down approach works as follows \cite{grotheer-etal-2020-covid}: first, classical NMF is performed on the whole data matrix $\textbf{X} \approx \textbf{WH}$, which extracts $r$ super-topics and outputs a dictionary matrix $\textbf{W}$ and coding matrix $\textbf{H}$. By setting a threshold in $\textbf{H}$, 
 a document can be classified into different topics and then form new data matrices $\textbf{X}_1, \textbf{X}_2, ..., \textbf{X}_r$, where the sub-matrix $\textbf{X}_i$ comprises all documents classified to a certain topic $i$. 
 On each sub-matrix, NMF is performed again to get detailed subtopics:
$$\textbf{X}_i\approx \textbf{W}_i\textbf{H}_i,$$ for $1\leq i \leq r$. This process can be applied multiple times in an iterative manner until no more coherent topics can be found. 

\subsection{Transformers and BERTopic}
A transformer is a neural network structure that depends on self-attention to relate outputs with inputs \cite{vaswani2017attention}. BERTopic is a topic modeling tool recently developed by Grootendorst \cite{grootendorst2022bertopic}. 
It incorporates the model architecture of Bidirectional Encoder Representations from Transformers (BERT), a pre-trained transformer-based model created by Google \cite{devlin-etal-2019-bert}, to produce topic representations from the sentences. Specifically, BERTopic employs the SBERT model \cite{reimers-2019-sentence-bert}, a variation of the BERT model that specializes in working on the sentence level, unlike the NMF approach, which discovers topics by transforming the entire document into a data matrix and decomposing it into two lower-ranked matrices. BERTopic then extracts topics by clustering on the sentence embedding space.

\begin{figure}[htbp]
  \centering
  \includegraphics[width=0.7\textwidth]{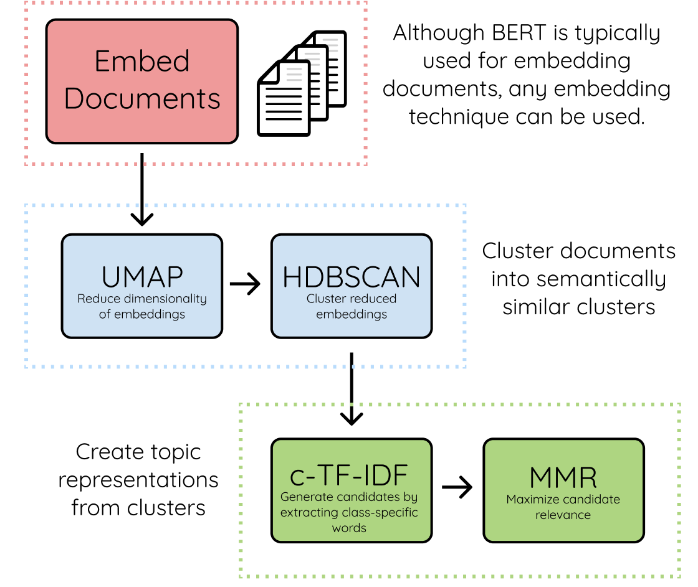} 
  \caption{The algorithm of BERTopic. \cite{grootendorst2022bertopic}}
\end{figure}
BERTopic consists of four major steps. It first uses the SBERT model \cite{reimers-2019-sentence-bert} to generate sentence embeddings from the transcripts. Through the use of SBERT, BERTopic can produce an intricate set of embeddings by accounting for the contexts of sentences in a document. However, the resulting embeddings are usually high-dimensional, making the clustering process difficult. Consequently, the second step comprises of 
BERTopic performing dimensionality reduction with UMAP \cite{mcinnes2018umap-software}. BERTopic 
then clusters these sentence embeddings through the hierarchical density-based clustering (HDBSCAN) algorithm \cite{mcinnes2017hdbscan}. HDBSCAN extends the DBSCAN clustering method, which clusters embeddings according to their densities and sparsities in the embedding space, to a hierarchical clustering method. It does so by constructing a spanning tree from the data and sorting neighboring edges of the tree to form hierarchies from the bottom to the top.

Since HDBSCAN detects noise in the embedding space, it also classifies some of the redundant or unrelated sentences as outliers. HDBSCAN then equates each cluster with a distinctive topic and extracts the words that are the most representative of such a topic. To quantitatively determine how representative a word is, HDBSCAN
computes the class-based term frequency-inverse document frequency (c-TF-IDF) score of each word in its topic. The c-TF-IDF score is given by the formula as used in \cite{grootendorst2022bertopic} 
\[ W_x = f_{x,c} \cdot \log \left(1+\frac{A}{f_x}\right), \]
where $W_x$ is the c-TF-IDF score of word $x$, $f_{x,c}$ is the term frequency of word $x$ in cluster $c$, $f_x$ is the term frequency of word $x$ across all clusters, and $A$ is the average number of words per cluster.
Unlike the classic TF-IDF \cite{Ramos_usingtf-idf} 
score, the c-TF-IDF score is computed at the cluster level rather than at the document level. After computing the scores, BERTopic then selects the words with the highest scores to represent each topic. Thus, this prepares the topics for human interpretations.

\section{Methods}
\label{sec:methods}


When reviewing transcribed interviews, we find that they often contain information at various scales. Additionally, there are usually a certain number of subtopics under each of the broad, easily-detected topics, which are worth further exploring. Interviews also contain some small topics that might not be directly related to the overall topic. 
Topics related to experiences shared by smaller groups of interviewees are often less obvious and may not be directly related to the broad topics. It is important, however, to not ignore their experiences. To find more detailed topics and cover less common experiences in the interviewees to a greater extent, we are compelled to combine NMF, a model useful for finding generally interpretable broad topics and guiding the hierarchical topic modeling structure, and BERTopic, which is useful for finding more detailed subtopics under the higher-level broad topics at the sentence level and potentially discovering semantically informed topics in sentences that might be missed by NMF. Therefore, we propose a new method for topic modeling: Multi-Scale Hybridized Topic Modeling (MSHTM), which is the combination of NMF and BERTopic, as summarized in Algorithm \ref{alg:one}. 


MSHTM follows a hierarchical
structure. It first employs NMF on the whole document at the first level to extract topic keywords. MSHTM then splits up the interview documents into sentence-level documents. In an interview dataset, a sentence-level document usually represents the interviewee's responses to a question, and it needs to be split apart based on timestamp and length if any response is too long. As a result, each sentence-level document contains one to five sentences. 
The model then transforms the sentence-level documents into new encoding matrices using the established NMF model at the first step. After that, it clusters the sentences to the most aligned topics based on the threshold in the encoding matrix. The threshold is a calculated distribution of the coefficients across all documents. The coefficients are shown in the rows of the matrix $\textbf{H}$, which represent the relevance of a specific topic for all the documents, as discussed in Section \ref{subsec:NMF}. MSHTM assigns each sentence-level document to a topic if the sentence's coefficient related to this topic exceeds the mean plus one standard deviation of all sentences' coefficients with respect to this topic. The threshold is not a set value, and it can be modified based on users' needs. For example, if we intend to be more strict about the relevance of the sentences assigned to each topic, we can modify the threshold to a higher value, such as mean plus 2 times the standard deviation. Similarly, we can lower the threshold if we prefer more lenient relevance. Then it applies BERTopic on the sentences under each of the broad topics to uncover hidden subtopics. Finally, it extracts the sentences that belong to each subtopic for further use. A flowchart that explains the steps of MSHTM is shown in Figure \ref{flowchart}. 

\tikzstyle{block} = [rectangle, draw, fill=green!30, 
text width=18em, text centered, rounded corners, minimum height=2.5em]
\tikzstyle{line} = [draw, -latex']

\begin{figure}[!h]

  \centering
  \hspace*{-10pt}
  \begin{tikzpicture}[node distance = 1.5cm, auto]
    \node [block] (NMF) {\small Perform NMF on whole interview documents};
    \node [block, below of = NMF] (split) {\small Split the interview documents into sentence documents};
    \node [block, below of  = split] (new encoding) {\small Transform sentence documents into encoding matrices };
     \node [block, below of  = new encoding] (assign sentences) {\small Assign sentence documents to NMF topics};
     \node [block, below of  = assign sentences] (BERTopic) {\small Perform BERTopic on sentences under each NMF topic};
     \node [block, below of = BERTopic] (assign sentences 2) {\small Obtain subtopics and clustered sentences};
     
     \path [line] (NMF) -- (split);
    \path [line] (split) -- (new encoding);
    \path [line] (new encoding) -- (assign sentences);
    \path [line] (assign sentences) -- (BERTopic);
    \path [line] (BERTopic) -- (assign sentences 2);
    \end{tikzpicture}
    \caption{MSHTM flowchart}
\label{flowchart}
\end{figure}
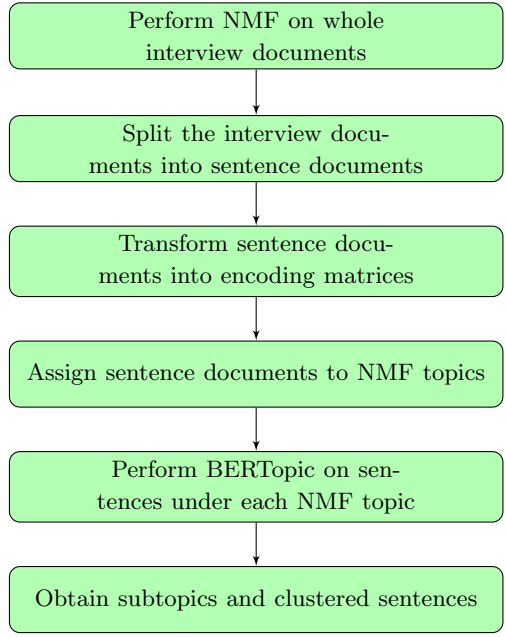

Our proposed MSHTM improves upon traditional approaches, being more flexible towards datasets with complex topic structures.
The hybrid nature of our method allows a sentence from the interview transcripts to be assigned to both a broad NMF topic as well as one or more BERTopic subtopic(s), which provides us with a better understanding of the topics discussed in a sentence. Moreover, the multi-scale nature of our method largely combines the advantages of both NMF and BERTopic. The topics extracted from MSHTM contain relatively accurate broader topics and detailed subtopics. The obtained subtopics also have better hierarchy and objectivity, compared to the traditional BERTopic algorithm. Additionally, MSHTM lowers the overall computational and storage cost 
of BERTopic. For instance, on a dataset with 450,000 sentences, MSHTM manages to run within 15 minutes, which is half the runtime of BERTopic on the same set. MSHTM also consumes around 8MB of RAM, which is more storage efficient than BERTopic as BERTopic consumes more than 12MB of RAM.

\RestyleAlgo{ruled}

\SetKwComment{Comment}{/* }{ */}

\begin{algorithm}[h]
\caption{Multi-Scale Hybridized Topic Modeling}\label{alg:one}
\KwIn{Data matrix $\textbf{X}$, sentence-level data matrix $\textbf{X}_{sentence}$, number of topics $k$}
$\textbf{W}, \textbf{H} \gets $ NMF($\textbf{X}, k$) \par
Transform the sentence-level documents $\textbf{X}_{sentence}$ into encoding $\textbf{H}_{sentence}$ using the existing NMF model and dictionary $\textbf{W}$\par
\For{each topic $i$}{
$\textbf{h}_i^T \gets$ extract the row vector from $\textbf{H}_{sentence}$ that corresponds to this topic 
\par
$hi\_mean, hi\_std \gets$ calculate the mean and standard deviation of all the entries in $\textbf{h}_i^T$ \par
\For{each sentence $j$}{
$\theta \gets$ $(\textbf{H}_{sentence})_{i,j}$, the $i$th topic coefficient for $j$th sentence \par
\If{$\theta\geq hi\_mean + hi\_std$}
{
    assign $j$th sentence to topic $i$\par
}
}
}

\While{topic index $i < k$}{
  perform BERTopic on all sentences under topic $i$\par
  extract the sentences belonging to each subtopic for later use\par
}
\end{algorithm}


\section{Datasets}
\label{sec:datasets}
In this section, we introduce the two datasets we use to test and demonstrate our method.  
We conduct experiments on both datasets to illustrate the performance and versatility of our proposed MSHTM. We compare the hierarchical structure and topic coherence of the generated results as well as the computational and storage cost of the methods. 

\subsection{USC Shoah Foundation’s
Visual History Archive English Transcripts}
This dataset consists of 984 English transcripts provided by the USC Shoah Foundation's Visual History Archive. Each transcript is a transcribed interview with a Holocaust survivor conducted by the Shoah Foundation. We  performed hierarchical topic modeling on this dataset using our novel model to detect both the collective and specific events that are brought up in the interviews, and the hidden hierarchies among these events. 

The transcripts all follow a similar question and answer structure. The total number of lines these transcripts contain ranges from 100 to 2100. The interquartile range (IQR) is from 330 to 550 lines. 
All the transcripts are stored in CSV file formats containing information on the file category, file number, identifier number for interviewer and interviewee, time stamp, and transcribed text. 

\begin{figure}[htbp]
  \centering
  \includegraphics[width=1\textwidth]{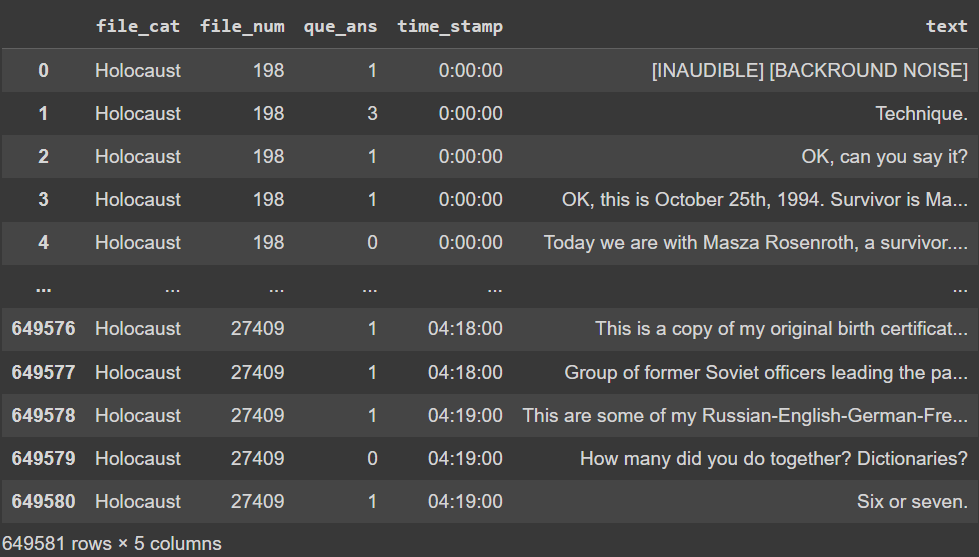} 
  \caption{The data frame that contains selected transcripts from USC Shoah Foundation’s
Visual History Archive English Transcripts.}
  \label{fig:data}
\end{figure}

Since some of this information is redundant or unrelated to our goal of extracting latent topics, we  clean the data extensively before implementing our proposed method. First, we combine all the transcripts into one data frame as shown in Figure \ref{fig:data}. Processing approximately 650,000 lines is a computationally expensive task. To reduce the number of lines we need to process, we only include the answers provided by the survivors and omit the questions asked by the interviewer. 
Importantly, we keep the \texttt{file\_num},  \texttt{time\_stamp}, and \texttt{text} columns from the transcripts in order to keep track of the original indexing. 

After cleaning the data in a top-level manner, we pre-process the fine-grained details. As shown in Figure \ref{fig:data}, some lines record non-verbal or non-transcribable instances such as `[INAUDIBLE]' and `[BACKGROUND NOISE]'. These instances are usually marked in capitals with square brackets. Furthermore, there are occasions when the transcriber is only able to guess the speaker's words. Similarly, lines under these occasions 
are also annotated with square brackets. Hence, we remove all these annotations from the data. We also pre-process the data by removing the commas in numbers, because the algorithms we are applying mistake these numbers as several different numbers which are separated by commas. For example, `3,000' will be interpreted as `3' and `000'.

Even after our cleaning process, the data still has inherent issues. One noteworthy challenge is handling grammatical mistakes in the testimonies. Since a lot of survivors are non-native English speakers, grammatical mistakes in their speech are unavoidable. 
Another issue is pronoun disambiguation. For example, the pronoun `they' can be referred to as the perpetrators, the victims, or the bystanders. These pronouns are at most times ambiguous especially when they are not given any context. 

\subsection{MediaSum}
\begin{figure}[htbp]
  \centering
  \includegraphics[width=0.9\textwidth]{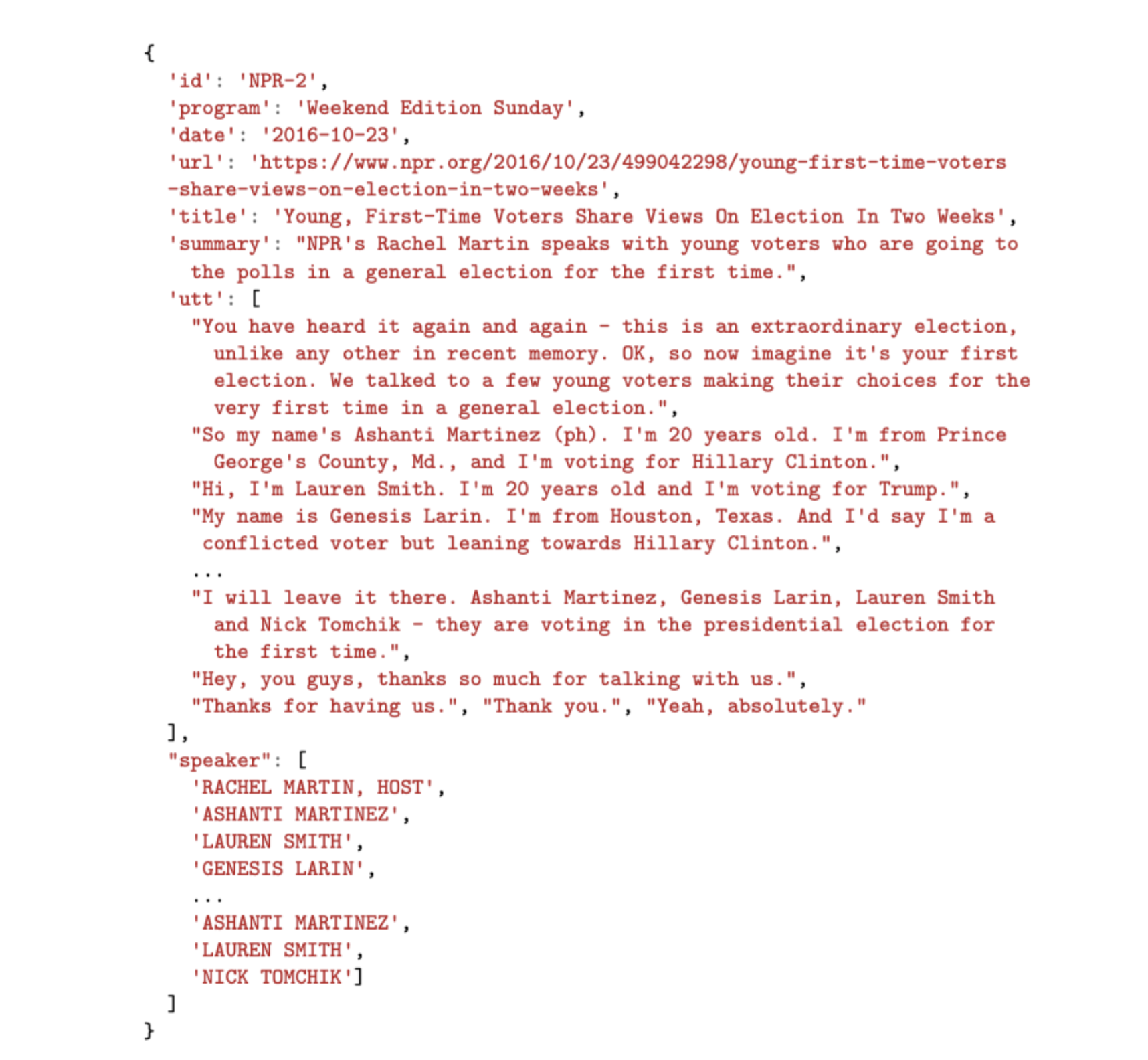} 
  \caption{MediaSum Data Example.}
  \label{fig:mediasum}
\end{figure}

MediaSum is a dataset containing 463.6K news interview transcripts from NPR and CNN \cite{zhu2021mediasum}. Compared to the USC Shoah Foundation’s
Visual History Archive English Transcripts, MediaSum dataset is much larger. There are more than 14 million lines in the MediaSum dataset. Hence, it poses a challenge to computational ability. This gives our method a chance to showcase its enhanced power in efficiency over BERTopic. The transcripts from MediaSum are also shorter in length. However, they cover a wider range of topics and thus contain a more diverse set of subtopics to discover. 
Before we use our methods on this dataset, we extract the \texttt{interview} and \texttt{id} data from the JSON files and remove some of the stop words to make the analysis easier. The list of stopwords is shown in \cref{stopwords_appendix}. 
An example of the MediaSum data is shown in Figure \ref{fig:mediasum}.

\section{Results and Visualizations}
\label{sec:results and visualizations}
In this section, we overview the results and visualizations obtained after applying classical NMF, BERTopic, and  MSHTM on the two datasets.

\subsection{Results for Classical NMF}
\label{Classical NMF}
In the following experiments, we construct a corpus matrix $\mathbf{X}$ where each column represents an individual interview. A bag-of-words TF-IDF embedding is adopted to represent each testimony (using parameters \texttt{max\_df}=0.8, \texttt{min\_df}=0.05 \texttt{ngram\_range}=(1,1)). TF-IDF determines word relevance by taking into account the term frequency of a word in a document and the inverse document frequency of the word across a set of documents \cite{Ramos_usingtf-idf}.
Any annotation content (e.g., `[NON-ENGLISH]' is removed, and a set of stopwords is also removed in the data cleaning process. For a more comprehensive list of stopwords, see \cref{stopwords_appendix}. 
\subsubsection{For USC Shoah Foundation’s
Visual History Archive English Transcripts}

Previously, M. Lee and T. Presner performed clustering on the questions asked in certain interviews\cite{sbertdashboard} to find out the general categories of the questions that have been given to interviewees. Inspired by the insights drawn from the process, we in turn perform topic modeling techniques on whole interview transcripts to gain an overview of all the topics discussed in interviews. We can then group interview contents into different topics, which provides clustered interview segments that can be used in further analysis. Below are the results we obtain from performing NMF on the whole transcript. 

In Table \ref{table:1}, 
we show the results from running classical NMF with eight 
coherent and interpretable topics. For each topic, the 15 most relevant keywords are displayed. For example, Topic 2 is related to Schindler's list, which refers to the list of Jews saved by Oskar Schindler, a German industrialist who provided a haven for approximately 1,200 Jews during the Holocaust by employing them in his enamelware and ammunition factories. Amon Goeth was the notorious SS officer who served as the commandant of the Krakow-Plaszow concentration camp, from where the Schindler Jews came. Note that the compound names ``Oskar Schindler,'' ``Amon Goeth,'' and ``Krakow Plaszow'' 
are broken down into single words due to the parameter \texttt{ngram\_range} being set to $(1,1)$ in the TF-IDF vectorizer. 
This could be avoided by adjusting the \texttt{ngram\_range}. However, this would create other issues like having slight variations of the same compound word (e.g., ``Krakow Plaszow'', ``Krakow'', and ``Plaszow Krakow'') appear as distinct words. Hence, we choose our \texttt{ngram\_range} to be $(1,1)$ to avoid these complications.

Topic 5 is related to the persecution of Jews in Hungary. 
Arrow refers to the Arrow Cross Party, which was a far-right Hungarian party that massacred thousands of Jews. Topic 7 is related to the infrastructure of concentration camps. Auschwitz-Birkenau was a concentration and extermination camp that held and executed millions of prisoners. Notably, the words ``gas" and ``crematorium" also appear in this topic. \\ 
\begin{table}[h!]
\centering
\scriptsize
\begin{tabular}{cccccccc}
 \hline
 Topic 1& Topic 2 &Topic 3 &Topic 4& Topic 5 &Topic 6 &Topic 7&Topic 8\\
Ghetto & Schindler &Post-War &Unkraine&Hungary&FamilyMemories&Auschwitz&Bergen-Belsen\\
 \hline
 ghetto & schindler & france & ukrainian & budapest & sort & auschwitz & holland \\
warsaw & plaszow & french & village & hungarian & actually & ss & amsterdam \\
Lodz & krakow & berlin & woods & hungary & guess & barrack & dutch \\
vilna & cracow & vienna & guy & hungarians & point & barracks & england \\
poles & ghetto & hitler & russia & arrived & wonderful & birkenau & belsen \\
treblinka & brunnlitz & ship & kill & labor & obviously & crematorium & london \\
apartment & factory & states & russians & romania & aunt & block & van \\
uprising & list & paris & till & arrow & uncle & gas & allowed \\
factory & oskar & united & forest & swedish & great & prisoners & bergen \\
lithuanian & goeth & british & ghetto & auschwitz & grandmother & soup & australia \\
czestochowa & gross & visa & gonna & actually & eventually & clothes & hiding \\
lithuania & rosen & american & kids & slovakia & recall & ghetto & underground \\
hospital & auschwitz & belgium & hiding & 1944 & difficult & transport & baby \\
flat & barracks & england & partisans & danube & having & factory & hospital \\
russians & apartment & nazis & farm & russians & saying & number & picked \\
 \hline
\end{tabular}
\caption{Top 15 keywords generated by NMF on all testimonies from USC Shoah Foundation’s
Visual History Archive English Transcripts.}
\label{table:1}
\end{table}

\subsubsection{For MediaSum dataset}

In Table \ref{table:newsdata12}, we show the results from running classical NMF with 12 topics, which give coherent and interpretable topics. Compared to the Shoah dataset, the MediaSum is a significantly larger dataset, and NMF can give a variety of topics in a relatively short amount of time. A brief discussion of the topics generated is as follows: domestic politics (Topic 1, Topic 2), elections (Topic 5, Topic 11), international affairs (Topic 3, Topic 7, Topic 8, Topic 12), police issues (Topic 9), economics (Topic 4), and weather (Topic 10). These help us uncover some of the most popular news topics over the past twenty years.
We named each of these topics based on the results on NMF.

Notably, we observe that ``Hillary Clinton'' appears in two topics, both Topic 5 and Topic 11. At first sight, it seems redundant. However, after some investigation, we find that Topic 5 is about the 2016 election and Topic 11 is about the 2008 election. This demonstrates the ability of NMF to differentiate various meanings of the same word.

\begin{table}[h!]
\centering
\footnotesize
\begin{tabular}{cccccc}
 \hline
 Topic 1& Topic 2 &Topic 3 &Topic 4& Topic 5& Topic 6 \\
 \hline
 president & court & russia & percent & trump & think \\
house & case & russian & market & donald & like \\
white & judge & putin & economy & president & yes  \\
democrats & trial & russians & money & campaign & really \\
republicans & attorney & intelligence & company & clinton & mean \\
senate & jury & syria & year & hillary & want \\
congress & justice & investigation & jobs & think & lot \\
administration & supreme & fbi & companies & said & time \\
think & law & information & tax & mean & said \\
said & defense & president & china & like & things \\
bush & legal & election & new & republican & thing \\
obama & evidence & meeting & business & saying & kind \\
committee & charges & isis & oil & michael & good \\
republican & federal & security & stock & cnn & way \\
senator & investigation & foreign & billion & media & tonight \\
 \hline
\end{tabular}
\begin{tabular}{cccccc}
 \hline
 Topic 7&Topic 8&Topic 9&Topic 10&Topic 11 & Topic 12\\
 \hline
korea & iraq & police & storm & clinton & israel \\
north& isis & officers & morning & campaign & iran \\
nuclear & iraqi & officer & water & hillary & minister  \\
south & al & shooting & area & obama & prime  \\
china & war & old & new & republican & peace \\
chinese & troops & gun & city & voters & nuclear \\
president & military & shot & weather & party & united \\
weapons & forces & family & miles & senator & government \\
military & baghdad & car & rain & mccain & deal \\
united & qaeda & killed & hour & vote & states \\
test & afghanistan & man & plane & democratic & world \\
states & syria & video & cnn & race & international \\
regime & bush & suspect & coast & election & east \\
talks & security & city & good & bush & syria \\
world & pentagon & year & flight & candidates & countries \\
\hline
\end{tabular}

\caption{Top 15 keywords generated by NMF on all news from MediaSum dataset.}
\label{table:newsdata12}
\end{table}

\subsection{Results for BERTopic}
Below are the results of BERTopic on the two datasets.
\subsubsection{For USC Shoah Foundation’s
Visual History Archive English Transcripts}
\begin{figure}[h]
  \includegraphics[width=1\textwidth]{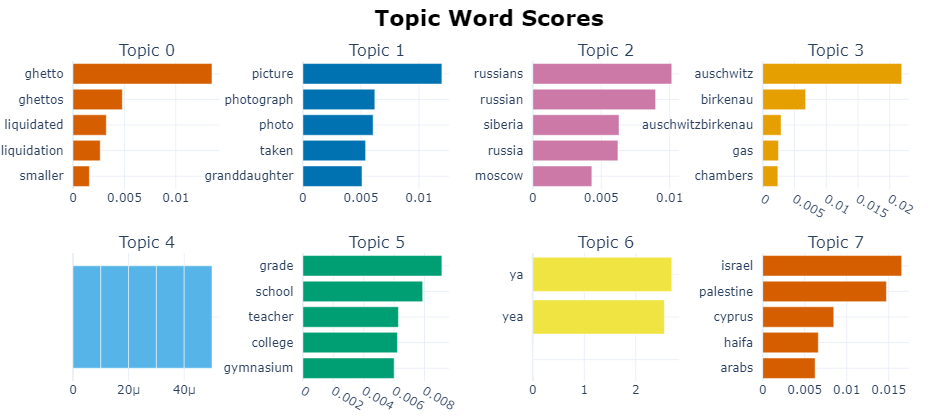} 
  \caption{The most representative words of the top 8 topics from USC Shoah Foundation’s
Visual History Archive English Transcripts.}
  \label{fig:barplot}
\end{figure}

After running BERTopic on all the 984 Shoah transcripts, BERTopic outputs approximately 2300 topics. Figure \ref{fig:barplot} shows the bar plots of the representative words weighed by their c-TF-IDF scores in the top eight topics. We infer that Topic 0 is about the liquidation of the ghetto since `ghetto' and `liquidated' are two of the most representative words. However, BERTopic also extracts some topics that reflect the structure of the interviews themselves. For instance, the existence of Topic 1 is because all the survivors were asked to show the interviewer a picture of their families and themselves. 

We notice that Topic 4 has abnormal results. This is a result of removing stop words by the CountVectorizer (detailed in \cref{stopwords_appendix}), as topic 4 comprises of sentences with a single `yes'. 

\begin{figure}[h]
  \includegraphics[width=1\textwidth]{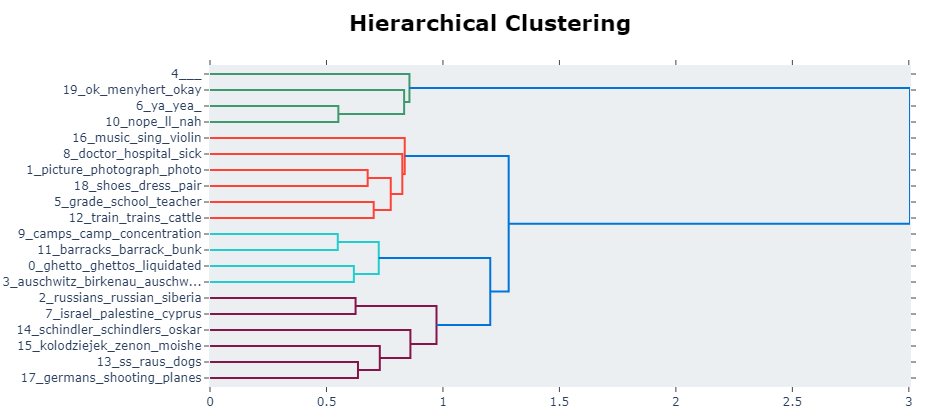} 
  \caption{The hierarchical map of the top 20 topics from USC Shoah Foundation’s
Visual History Archive English Transcripts.}
  \label{fig:hiermap}
\end{figure}
Figure \ref{fig:hiermap} shows the hierarchy of the top 20 topics generated by BERTopic. 
In particular, the 20 topics are grouped into four main clusters as shown in their colors. For instance, the green cluster contains all the topics that are associated with `yes' and `no'. The blue cluster encapsulates all topics relating to concentration camps and ghettos. Looking closer into this cluster, Topics 9 and 11 are mapped to the lowest level due to the correlation between both topics. More precisely, camps and barracks are strongly related because prisoners of concentration camps are kept in the barracks.

\subsubsection{For MediaSum dataset}
\label{sec:BERTonMedia}
Although BERTopic is a powerful tool for topic modeling, its intricate mechanism makes it expensive storage-wise. In particular, BERTopic requires a lot of memories when 
running on large datasets. Since the MediaSum dataset contains more than 14 million sentences, BERTopic aborts the program as it is too memory intensive. Hence, we cannot run BERTopic on the entire MediaSum dataset. Thus, to get some preliminary results from running BERTopic on the dataset, we randomly select a subset with a sample size of 1,000,000.  

\begin{figure}[htbp]
  \includegraphics[width=1\textwidth]{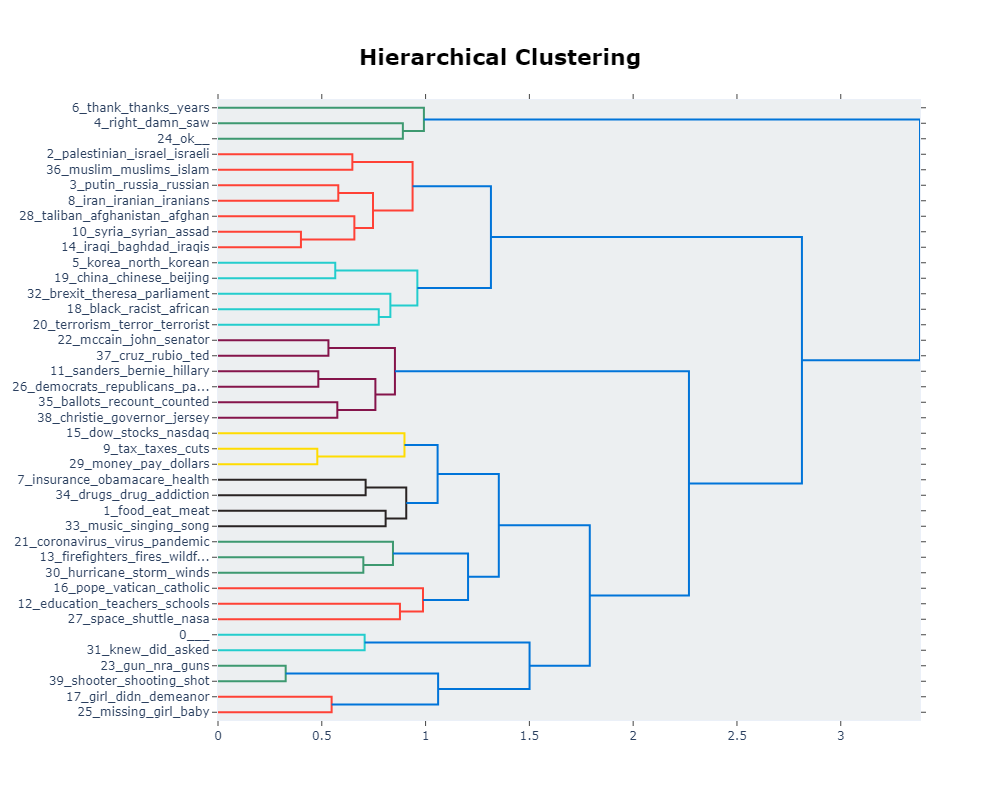} 
  \caption{The hierarchical map of the top 40 topics from running BERTopic on the sample from MediaSum dataset.}
  \label{fig:bert_mediasum}
\end{figure}

As seen in Figure \ref{fig:bert_mediasum}, BERTopic manages to retrieve similar topics like politics (with the democrats vs republicans and the Hilary Clinton topics) and foreign relations (with the China and Noth Korea topics) as discovered by NMF in Table \ref{table:newsdata12}. However, it also fails to include certain topics like Donald Trump and court topics, which are prominent topics found by NMF in Table \ref{table:newsdata12}. This is because we are running on a sample of one million sentences as opposed to the original size of 14 million. Hence, the result in Figure \ref{fig:bert_mediasum} is not the best in reflecting the actual prominent topics of the dataset. To solve this issue, we will instead use MSHTM on the dataset.

\subsection{Results for MSHTM}
\label{sec:MSHTM results}
We show 
some example results for our MSHTM method using the two datasets we introduced before. We demonstrate that this hybridized model generates topics in a clear hierarchical structure and notably reduces the memory usage and run time of BERTopic. 
Since the first layer of NMF dissects the documents into smaller portions, this increases the efficiency of the BERTopic step both with respect to memory and running time. 
Therefore, our hybridized model is a powerful tool that embraces the complementary relationship between NMF and BERTopic.

\subsubsection{For USC Shoah Foundation’s
Visual History Archive English Transcripts}

We have applied our proposed hybrid model with eight top-level topics to the transcripts. Referring to the top-level topics obtained from classical NMF Table \ref{table:1} in Section \ref{Classical NMF},
the number of subtopics under these top-level topics ranges from around 25 to 150, with the Schindler topic and France topic generating the smallest and largest numbers of subtopics respectively. 

\begin{figure}[htbp]
  \includegraphics[width=1\textwidth]{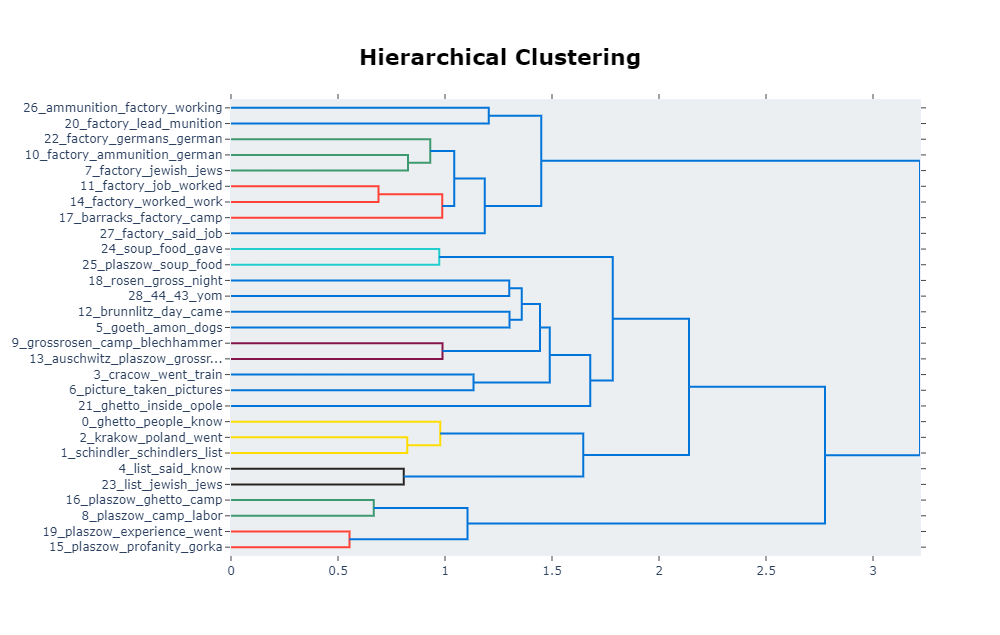} 
  \caption{The hierarchical map of the subtopics under the Schindler top-level topic from USC Shoah Foundation’s
Visual History Archive English Transcripts.}
  \label{fig:hybrid_result}
\end{figure}
 
Here is a closer look into the Schindler subtopics in Figure \ref{fig:hybrid_result}. Notice that the subtopics on the top of the map are under one higher-level topic that associates with the factory that Oskar Schindler oversaw. The remaining topics on the bottom half are grouped into another high-level topic that is mostly related to concentration camps and ghettos.

One of the notable benefits of our proposed hybridized model is that it can assign one line to multiple high-level topics. Take the following quote as an example:
\begin{quote}
    `1944, the Germans came, and they round up the ghetto and send everybody out. We didn’t know, it was a terrible confusion. We didn’t know where we were going. But they sent they round them up and they sent them to Auschwitz. They closed the Lodz ghetto.'
\end{quote}
The hybridized model successfully classifies this quote into two top-level topics: the Auschwitz topic and the ghetto topic, whereas using BERTopic can only assign one topic to each sentence. This ability of multiple topic assignments is inherited from the flexibility of NMF. Within each top-level topic, with the aid of BERTopic, the quote goes into a subtopic under the corresponding top-level topic. In particular, the quote is assigned to a subtopic associated with `ghetto' and `transport' and another subtopic associated with `Jews' and `Germans', both under the Auschwitz and ghetto topics respectively. Given the size of each sentence-level document, it will be more difficult to obtain topics at this fine-grained level if only NMF is used. Although 
the NMF layer manages to identify this quote with the main themes like `Auschwitz' and `ghetto', it fails to capture the finer-grained details like the victims and the perpetrators of the event, which are shown in the subtopic related to `Jews' and `Germans' generated by the hybrid model. On the other hand, if we only use BERTopic, we will not get the ability of multiple topic assignments. This might be problematic for sentences that contain more than one theme. Hence, this shows the power of our hybridized model in assigning multiple topics without missing the finer grains. 

 
\subsubsection{For MediaSum dataset}
For MediaSum dataset, we apply our proposed hybridized model with 12 top-level topics to the interviews. The number of subtopics under these 12 top-level topics ranges from around 1000 to 4000.

\begin{figure}[htbp]
  \includegraphics[width=1\textwidth]{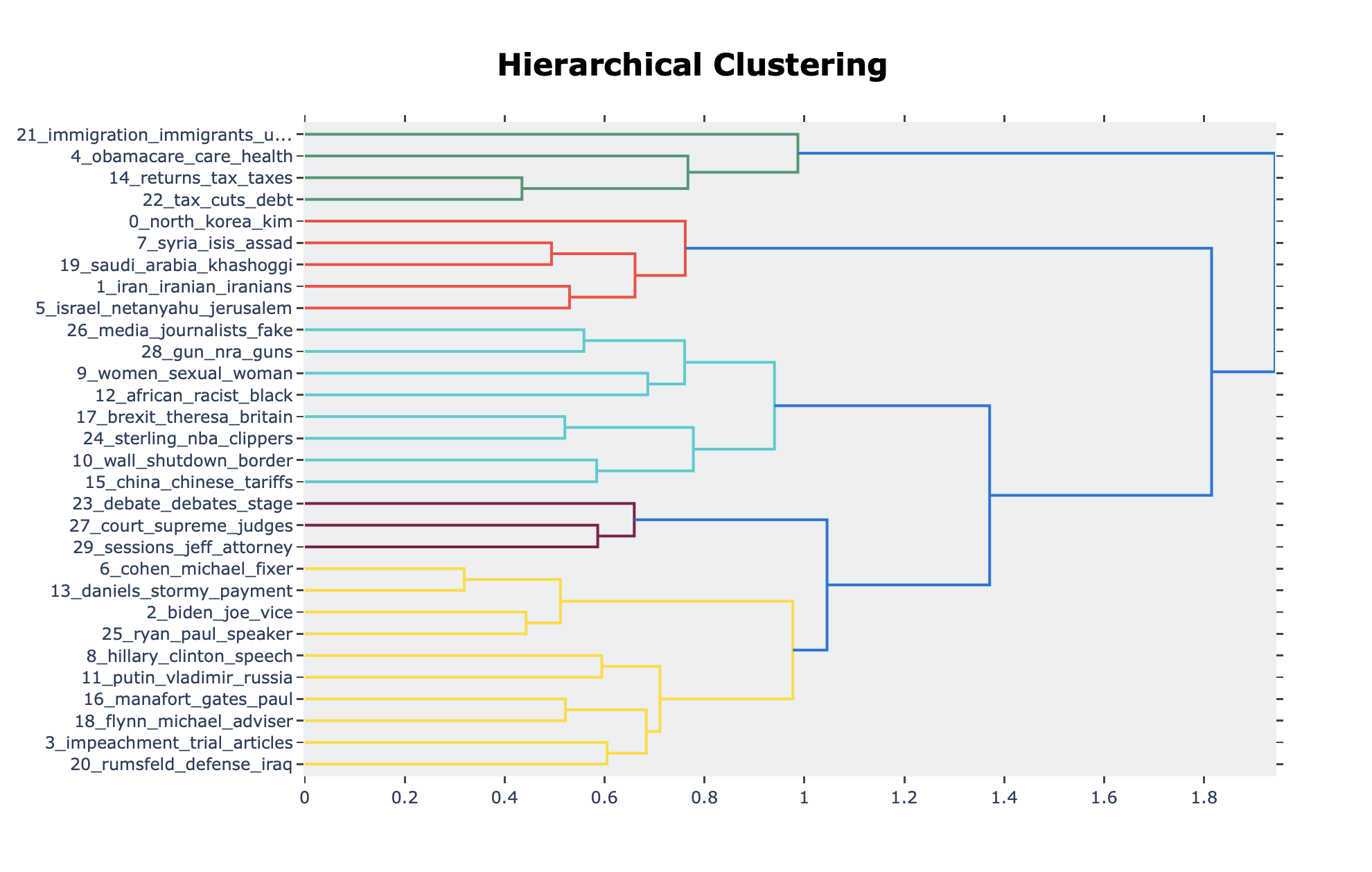} 
  \caption{The hierarchical map of the subtopics under the presidential top-level topic from MediaSum dataset.}
  \label{fig:hybrid_result2}
\end{figure}
 Figure \ref{fig:hybrid_result2} shows the subtopics within the presidential topic. Notice that the subtopics on the top of the map (in red) are under one higher-level topic that associates with countries in the Middle East and Asia. The topics in the middle (in cyan) represent various political problems. The topics on the bottom (in yellow) are grouped into another high-level topic that is related to a collection of notable political figures. We also note that, when dealing with large-scale datasets like MediaSum, it is also helpful to perform Hierarchical NMF as the first layer in MSHTM. This optional HNMF layer can lead to even lower computational costs and clearer hierarchical structures.


\section{Discussions}
\label{sec:discussions}
In addition to the benefit discussed in \ref{sec:MSHTM results}, which shows that MSHTM can extract more comprehensive and detailed topics from sentences, another benefit of MSHTM is the lower computational cost. One aspect is its shorter run time. Due to the intricacy of BERTopic, BERTopic takes 30 minutes to run all the Shoah transcripts. On the other hand, our hybridized model manages to process the complete dataset in just 15 minutes. The shorter run time is an outcome of running and reusing the same BERTopic model on all the smaller subsets created by NMF. Another aspect is its lower storage cost. Recalling Section \ref{sec:BERTonMedia}, BERTopic fails to run on the entirety of MediaSum dataset due to the intense memory usage. On the other hand, the hybridized model relieves memory consumption by breaking the dataset into smaller portions through NMF and running BERTopic on all portions without the need of reconstructing the BERTopic model every time. This allows the model to run on a larger subset (or even the entire dataset) while utilizing the power of BERTopic.

Overall, as shown in Section \ref{sec:results and visualizations}, MSHTM is helpful to understand and organize large amounts of unstructured corpora, such as the Shoah Transcripts and MediaSum dataset we used. For a study like this, this structure can help a user conveniently locate an event and access sentences under specific subtopics. 
From there, we can detect the shared experiences between different interviewees and their perspectives on the same topic, which may provide us with a more comprehensive and objective understanding of the said topic. We can also analyze the mainly involved agencies and actions for different topics and identify the roles they played in the evolution and the outcomes of the events.

\section{Conclusions}
\label{sec:conclusions}
We propose MSHTM to organize unstructured interview transcripts into meaningful groups and form meaningful interview transcript indexing without laborious human labeling, with more detailed classification and lower computational cost. The topics generated by the MSHTM are coherent, easily interpreted, and of a clear hierarchical relation. Moreover, the method does not overlook opinions and experiences described by a minority. The hidden topics found 
are valuable to analyze and gain new perspectives from the interview transcripts. They can be informative and educational to the general public, and create an organized structure for researchers to study interviews.

\section*{Acknowledgments}
We would like to thank Professor Deanna Needell for heading the project. We want to thank Professor Todd Presner, Leo Fan, and Michelle Lee for providing insights into their past work on  Holocaust survivor testimonies. We acknowledge the USC Shoah Foundation Visual History Archive for providing the transcripts and thank Professor Todd Presner for directing the UCLA Holocaust Digital Humanities research lab. Finally, we also want to give our deepest gratitude to Professor Michael Perlmutter, Professor Michael Lindstrom, and Joyce Chew for mentoring and guiding us throughout the project.

%
\bibliographystyle{plain}
\bibliography{references}

\appendix
\section{Appendix} 
\subsection{Stopword List}
\label{stopwords_appendix}
The list of stopwords we use for Shoah Foundation English Transcript Data (NMF) is the ENGLISH\_STOP\_WORDS from the sklearn package \cite{scikit-learn}, plus [``um'' and ``uh''].

The list of stopwords we use for MediaSum Data (NMF) is the ENGLISH\_STOP\_WORDS from the sklearn package \cite{scikit-learn}, plus [``s'', ``t'', ``don'', ``ve'', ``did'', ``got''].

\subsection{Discussion of Sparsity in NMF}
\begin{figure}[!ht]
     \centering
     \begin{subfigure}[b]{0.8\textwidth}
         \centering
         \includegraphics[width=\textwidth]{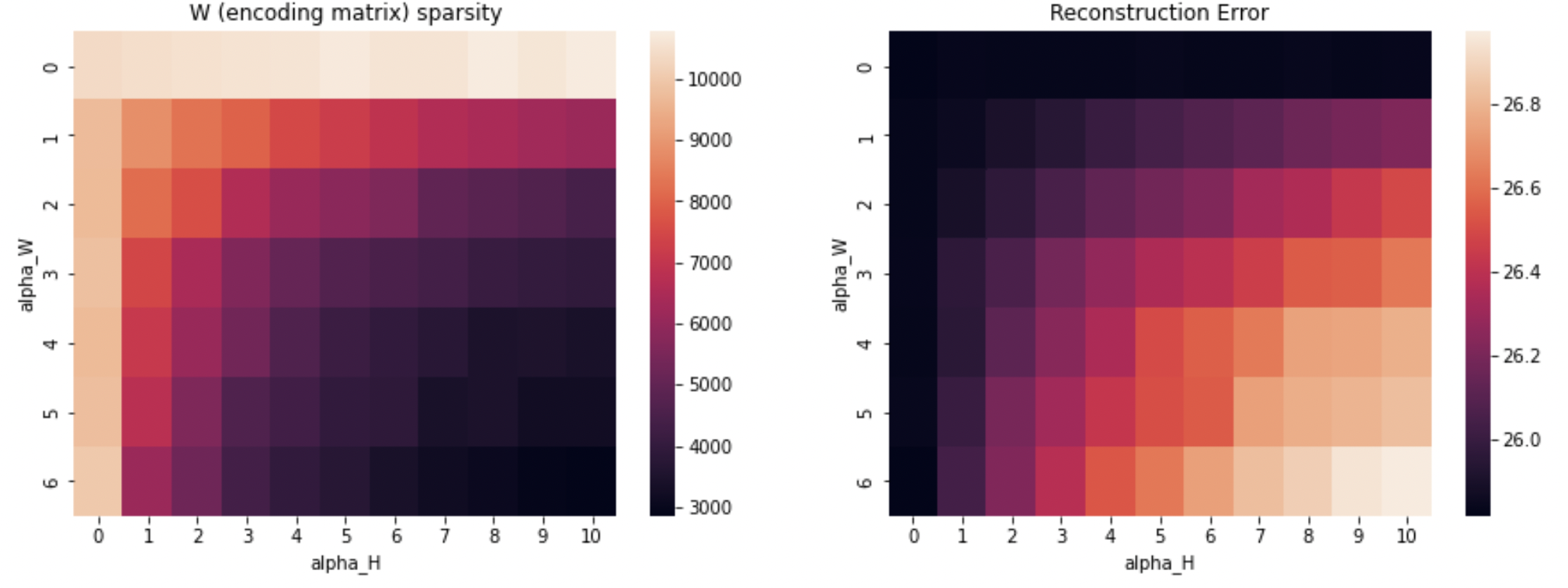}
         \caption{Varying alphas for both dictionary and coding matrix.}
         \label{fig:bothalpha}
     \end{subfigure}
     \hfill
     \begin{subfigure}[b]{0.8\textwidth}
         \centering
         \includegraphics[width=\textwidth]{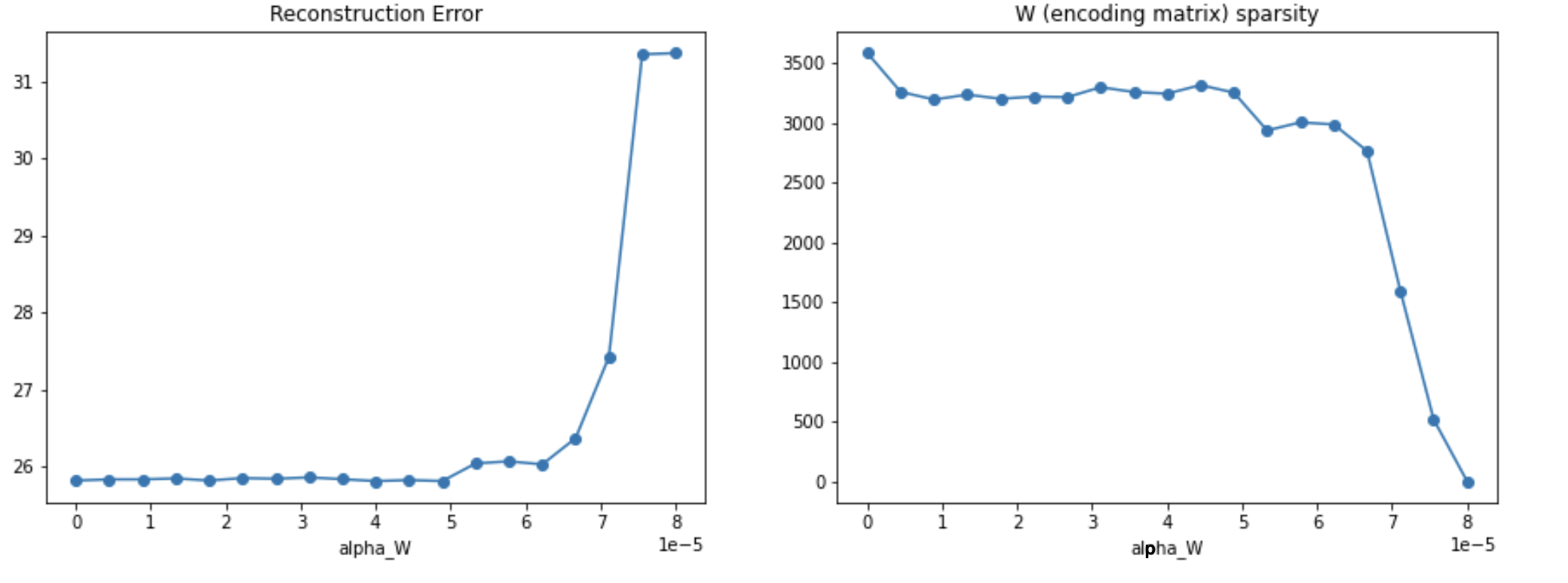}
         \caption{Varying alpha for coding matrix.}
         \label{fig:alphaw}
     \end{subfigure}
     
        \caption{Tuning L2 regularization parameters}
        \label{fig:sparsity}
\end{figure}

To increase the chance to classify the sentences into the correct topics, it is desirable to have the coding matrix $H$ of NMF to make it sparse. A common approach is to use L2 regularization. We can control that by tuning the parameter $alpha\_W$ and $alpha\_H$ in the NMF function from the sklearn package \cite{scikit-learn}. In Figure \ref{fig:sparsity}, we display the results obtained by varying $alpha\_W$ between $[0, 0.00006]$ and $alpha\_H$ between $[0, 0.0001]$.

\subsection{Normalized Point-wise Mutual Information}
Normalized Point-wise Mutual Information (NPMI) \cite{newman_2010} is a numerical metric to evaluate topic coherence after topic modeling. The formula can be seen below:

\[ \textbf{PMI} \left(x_i, x_j \right) = \log \frac{p \left (x_i, x_j \right)}{p \left (x_i \right)p \left (x_j \right)} ,\]
\[ \textbf{NPMI} \left(x_i, x_j \right) = \frac{ \textbf{PMI} \left (x_i, x_j \right)}{- \log p \left (x_i, x_j \right)} ,\]
where $x_i$ and $x_j$ are a pair of representative words taken from a topic. However, since this metric evaluates the topic coherence by only measuring the co-occurrence of the word pair, some of the topics with a high NPMI score do not seem coherent after human inspection. For instance, one of the topics with representative words such as `seven', `jaguars', and `mitts' has an approximately 0.65 NPMI score even though we cannot find any strong correlation among these words.

\end{document}